\documentstyle[preprint,aps]{revtex}

\begin{document}

\draft


\title{A Self-Consistent Microscopic Theory of Surface Superconductivity}

\author{Robert J. Troy$^*$ and Alan T. Dorsey}

\address{Department of Physics,  University of Virginia,\\
	McCormick Road, Charlottesville, Virginia 22901}

\date{\today}

\maketitle

\begin{abstract}

The electronic structure of the superconducting surface sheath in a
type-II superconductor in magnetic fields $H_{c2}<H<H_{c3}$ is
calculated self-consistently using the Bogoliubov-de Gennes equations.
We find that the pair potential $\Delta(x)$ exhibits pronounced
Friedel oscillations near the surface, in marked contrast with the results of
Ginzburg-Landau theory.  The local density of
states near the surface shows a significant depletion near the
Fermi energy due to the development of local superconducting order.
We suggest that this structure could be unveiled by
scanning-tunneling microscopy studies performed near the edge of a
superconducting sample.

\end{abstract}



The study of surface superconductivity was initiated some thirty
years ago by the seminal work of Saint-James and de Gennes \cite{saintjames63},
who predicted that a magnetic
field parallel to a superconductor-vacuum interface would nucleate
a superconducting ``surface sheath'' before the onset of superconductivity
in the bulk of the material.  By solving the linearized Ginzburg-Landau
(GL) equations subject to the boundary condition that the {\it normal
derivative}
of the order parameter vanish at the interface, they  found an
approximately Gaussian order parameter profile localized within a
coherence length $\xi$ of the interface;
the critical field for this surface superconductivity is $H_{c3}=1.69\,H_{c2}$,
where $H_{c2}$ is the bulk critical field for the Abrikosov flux-lattice
phase \cite{saintjames63,degennes66,saintjames69}.
This phenomenon has been confirmed by measurements
which observe a vanishing surface resistance at fields above $H_{c2}$
(for a review of the early experiments,
see Refs. \cite{degennes66}and \cite{saintjames69}).
Due to the presence of the superconducting nucleus there is also
a depletion of states near the Fermi energy, which is reflected in
a suppression of the tunneling conductivity at low bias, an effect
which has been observed in certain Pb-Bi and Sn-In alloys
\cite{guyon65}.

While the theory of surface superconductivity is quite complete within
the framework of GL theory, there remain several interesting unanswered
questions and problems which can only be addressed within a microscopic theory.
(1) In a microscopic theory the natural boundary condition for
the quasiparticle wavefunctions is that they vanish at the
superconductor-vacuum interface, so that superconducting pair potential also
vanishes at the interface.   If we naively apply this microscopic
boundary condition to the macroscopic GL equations, and solve
the linearized GL equations subject to the boundary condition that the
{\it order parameter} vanish at the interface (rather than its
{\it derivative}) we find $H_{c3}=H_{c2}$; i.e., there is no surface
superconductivity. However, we can not go back and infer that the microscopic
equations do not possess surface superconducting solutions; by the same token,
it cannot be taken for granted that the surface-superconducting solutions
of GL theory (with the zero derivative boundary condition) prove the existence
of
surface-superconducting solutions of the microscopic theory.  The two
approaches,
valid at different length scales, utilize different boundary conditions, and it
is generally difficult to connect the two \cite{stojkovic93}.
Whether surface superconductivity
exists within a realistic microscopic model of superconductors remains
an open question.
(2) The GL equations are derived using the {\it quasiclassical phase
approximation}, which neglects the effects of Landau level quantization
of the electronic states, and is valid in low fields.
In very high magnetic fields the Landau
level quantization can become important. Recent theoretical work
has predicted de Haas-van Alphen oscillations
in $H_{c2}(T)$, as well as possible re-entrant superconductivity
in very high magnetic fields \cite{rasolt92}.
Might there be re-entrant surface superconductivity which precedes the
re-entrant bulk superconductivity?  Again, the answer would
require a microscopic calculation which does not invoke the quasiclassical
phase approximation. (3) On a parallel note, Landau level quantization
in the presence of a surface produces
magnetic edge states, which have been the subject of intensive study in the
context of the integer and fractional quantum Hall effects \cite{stone}.
Understanding the role that edge states play in surface superconductivity
may help us in answering the basic question: Why is superconductivity
favored at a surface?
(4) In the mixed state of type-II superconductors the spatial variation of the
pair potential can produce a rich structure in the
local density of states (LDOS) \cite{shore89,gygi}, which can be measured
directly using a scanning-tunneling microscope (STM) \cite{hess}.
Analogous structure in the LDOS will be produced by the superconducting
surface sheath, which should be observable using a STM.  Such STM studies
would provide the first direct image of the surface sheath, and provide
valuable information about the pair potential profile and {\it local}
electronic structure;  all of the previous
experimental studies have only explored {\it averaged} properties, such
as the pair potential averaged over the sample
\cite{degennes66,saintjames69,guyon65}.   We note that previous attempts
at a microscopic theory of surface superconductivity
have been confined to analytical \cite{abrikosov65,kulik69,hu69}
or numerical \cite{scotto92} solutions of the linearized gap equation.
All of these works invoke the quasiclassical phase approximation,  and make
approximations which are equivalent to assuming that the derivative of the
pair potential vanishes at the surface; they therefore do not address the
issues
which we have raised above.  There have been no calculations of the LDOS.

In this Letter we will discuss our first attempts at addressing some of the
questions raised above by numerically solving the
the microscopic Bogoliubov-de Gennes (BdG) equations\cite{degennes66}
self-consistently, in a magnetic field with realistic boundary conditions
at the superconductor-vacuum interface.  We retain fully the Landau level
(and edge state) structure.   To make the calculations computationally
tractable we assume a two-dimensional geometry, which is also a good
approximation for many layered superconductors (see below).
Our results show that the microscopic BdG equations do indeed admit a
superconducting solution localized near the surface, for fields $H>H_{c2}$.
The pair potential
vanishes at the surface, but rises rapidly and eventually looks like the
GL solution; there is a narrow ``boundary layer'' near the surface in
which the GL solutions break down.  However, unlike the GL solution we find
large amplitude Friedel-like oscillations in the pair potential.
Our LDOS calculations
show a suppression at low energies, in the regions where
the pair potential is a maximum; it should be possible to resolve this
structure in an STM measurement.  The remainder of this paper is
organized as follows:  Following a brief review of the BdG formalism,
we will discuss our numerical methods.
We will then present our results for
the self-consistent pair potential and the LDOS, for a
particular choice of parameters.   Further details of these calculations
will appear in Ref.\ \cite{troy}.

\paragraph*{The Bogoliubov-de Gennes Equations.}
The BdG equations\cite{degennes66} for the quasiparticle amplitudes
$u_{i}({\bf r})$ and $v_{i}({\bf r})$ with excitation energy $\epsilon_{i}>0$
(measured relative to the Fermi energy) are
\begin{equation}
\label{bdg}
\pmatrix{{\cal H}_{e} & \Delta({\bf r})\cr
 \Delta^{*}({\bf r}) &  -{\cal H}_{e}^{*} \cr}
 \pmatrix{u_{i}({\bf r})\cr v_{i}({\bf r})\cr}
=\epsilon_{i}\pmatrix{u_{i}({\bf r})\cr v_{i}({\bf r})\cr},
\end{equation}
with $\Delta({\bf r})$  the pair potential.
The single-particle electron Hamiltonian is
\begin{equation}
   {\cal H}_{e} = {1\over{2m^*}}\left[ -i\hbar\nabla
     - {e\over{c}}{\bf A}({\bf r})\right]^{2} + V({\bf r}) - E_{F},
\end{equation}
where $E_{F}$ is the Fermi energy, $V({\bf r})$ is the surface potential,
and ${\bf A}({\bf r})$ is the vector potential
(we do not consider any effects of spin).  Any effects of the band structure
of the material are subsumed in the effective mass $m^{*}$.
The pair potential must be determined
self-consistently from the solutions of the BdG equations, as
\begin{equation}
\Delta({\bf r}) = g\sum_{\epsilon_{i}\leq \hbar\omega_{D}}
     v_{i}^{*}({\bf r})u_{i}({\bf r})
     \lbrack 1-2f(\epsilon_{i})\rbrack,
\label{order}
\end{equation}
where $g$ is the BCS attractive coupling, $\omega_{D}$ is the Debye
frequency, and $f(\epsilon)$ is the Fermi function.
The vector potential must also be determined self-consistently
using Ampere's law with the current density determined by the quasiparticle
wavefunctions \cite{degennes66,gygi}.  Once the quasiparticle wavefunctions
have been computed self-consistently, we can calculate the thermally broadened
local density of states,
\begin{eqnarray}
N({\bf r},E) & =&-  \sum_{i} \left[ |u_{i}({\bf r})|^{2} f'(E-\epsilon_{i})
                                 \right.  \nonumber \\
      &  &\qquad\qquad  + \left. |v_{i}({\bf r})|^{2}
f'(E+\epsilon_{i})\right],
\label{ldos}
\end{eqnarray}
with $f'(\epsilon)= \partial f/\partial \epsilon$.
This quantity is proportional to the local differential tunneling conductance
which is measured in a STM experiment.

We now take the magnetic field ${\bf H}= H\hat{z}$ to be parallel
to the vacuum/superconductor interface at $x=0$, with the superconductor
occupying the half-space $x>0$.   As we will eventually be interested in
modeling
quasi-two dimensional materials, we will neglect dispersion in the
$z$-direction.  Assuming the interface to be perfectly
impenetrable, we have $u(x=0)=v(x=0)=0$ (and therefore $\Delta(x=0)=0$).
In the Landau gauge
${\bf A} = (0,Hx,0)$ the BdG equations are translationally invariant
in the $y$-direction, and so we factor out this dependence as
\begin{equation}
\Delta({\bf r}) = e^{i 2X_{0}y/l^{2}} \Delta (x),
\label{Delta}
\end{equation}
\begin{equation}
\pmatrix{ u({\bf r}) \cr v({\bf r}) \cr}  = e^{i (x_{0}\pm X_{0}) y/l^{2}}
   \pmatrix{ u_{x_{0},n}(x) \cr v_{x_{0},n}(x) \cr}.
\label{uv}
\end{equation}
Here $l^2 = \hbar c/e H$ is the magnetic length, $X_{0}$ the orbit center
for the pair potential, $-\infty<x_{0}<\infty$  the orbit center
for $u$ ($-x_{0}$ is the orbit center for $v$), and $n=0,1,2,\ldots$ is a
Landau-level index which counts the number of
nodes of the wavefunctions; the sums in Eqs. (\ref{order}) and (\ref{ldos})
are over all $(x_{0},n)$.  We are left with a set of coupled one-dimensional
equations for $ u_{x_{0},n}(x)$ and $ v_{x_{0},n}(x)$.   Because of
our boundary condition at $x=0$ the corresponding
eigenvalues $\epsilon_{n}(x_{0},X_{0})$ will depend upon the positions
of the orbit centers, unlike the bulk case in which the energies are
degenerate with respect to $x_{0}$.
The effort involved in solving these equations can be substantially
reduced by finding both positive and negative energy solutions for $x_{0}>0$
and taking advantage of the transformation \cite{degennes66,gygi}
$\epsilon\rightarrow -\epsilon$,$u({\bf r})\rightarrow v^*({\bf r})$,
$v({\bf r})\rightarrow- u^*({\bf r})$, to convert the negative energy
solutions for $x_{0}>0$ into positive energy solutions for $x_{0}<0$.

\paragraph*{Method of solution.}  The BdG equations are solved iteratively,
as follows.  We start with an initial guess for the amplitude and
the phase of the pair potential, taken from GL theory, for instance.
We then fix the orbit center $x_{0}$ and calculate the wavefunctions
and energies  in the range $0<\epsilon <\hbar\omega_{D}$,
by writing the BdG equations as a set of finite-difference equations.
The lattice spacing is determined so that variations on the
scale of $\pi/k_{F}$ can be resolved.
The resulting matrix equations are sparse, and can be diagonalized
using standard packages (we use LAPACK).
This process is then repeated for new values of
$x_{0}$.  The {\it range} of values of $x_{0}$ is determined so that the
highest energy states (of energy $\hbar\omega_{D}$) are approaching their bulk
behavior, i.e., becoming independent of $x_{0}$, which occurs
at  $x_{0} = l [ 2 ( E_{F} + \hbar \omega_{D})/\hbar \omega_{c}]^{1/2}$, with
$\hbar\omega_{c} = \hbar e H /m^{*}c$ the cyclotron energy.
The {\it spacing} between these
points is again determined by requiring that structure on the scale of
$\pi/k_{F}$ can be resolved.  Once all of the wavefunctions have been
determined, the amplitude of the pair potential is recalculated
from Eq.\ (\ref{order}) by summing over $x_{0}$ and $n$.  The
phase of the pair potential is also recalculated by using the
self-consistency condition for the vector potential\cite{degennes66,gygi}.
The entire process is then repeated until the relative error in the
order parameter between successive iterations is less than $0.02$.

Several cases were tested to determine the reliability of the algorithm.
When $\Delta({\bf r})=0$ we have reproduced the wavefunctions and
spectrum for electrons in a constant magnetic field in the presence
of an impenetrable surface \cite{spectrum}.
The eigenvalues for states with orbit centers at the surface
($x_{0}=0,\ X_{0}=0$)
were within $1\% $ of those found analytically;  for $x_{0}$ large
the usual bulk Landau levels were obtained.
We have also used several different initial guesses for the pair
potential, including the GL form and a constant pair potential.
The final results are insensitive the form of the initial
pair potential, but convergence is expectedly slower for the constant
pair potential.
In the results shown here we have used a Gaussian profile
centered near the surface $x=0$ for the initial pair potential amplitude,
and we have used for our initial $X_{0}$ the value obtained in
GL theory \cite{degennes66}.

We have chosen to model a layered (i.e., two dimensional) material
whose parameters obey the weak-coupling BCS relations.
We took the Fermi surface to be cylindrical with $m^{*}_{x-y} \ll m^{*}_{z}$
and $m^{*}_{x-y} = 2 m$, where $m$ is the electron mass.
Assuming the zero temperature gap $\Delta(0) = 1.1$ meV and the
zero temperature coherence length $\xi(0)=100\ {\rm \AA}$
yields a Fermi velocity of
$v_{F} = \pi \Delta(0)\xi(0)/\hbar = 5.27 \times 10^6\ {\rm cm/s}$,
$k_{F}^{-1} = 11.0\ {\rm \AA}$,  and
$E_{F} = 15.8\ {\rm meV}$.
With $\hbar\omega_{D} = 15\ {\rm meV}$ and $gN(0)=0.30$,
the zero field critical temperature is $T_{c}=7.25\ {\rm K}$, and the
zero temperature quasiclassical critical field is
$H_{c2}(0) = 0.722\, \phi_{0}/2\pi\xi^{2}(0)= 2.38\ {\rm T}$ \cite{hc2}.
These parameters are similar to those used by
Gygi and Schl\"uter in their study of the core structure of vortices in
${\rm NbSe}_{2}$ \cite{gygi}, which showed good agreement with
STM measurements \cite{hess}.
Our Fermi energy is probably unrealistically low; it was
chosen to make our computations tractable, and thus represents a compromise
between numerical efficiency and realistic modeling.   We do not expect
any qualitative changes in our results for larger values of $E_{F}$.

We have chosen to present results for a temperature of 2 K ($T/T_{c} = 0.28$),
and a  magnetic field of 4 T ($H/H_{c2}(T) = 1.83$), so that
$l=128\ {\rm \AA}$ and $\hbar\omega_{c}= 0.23\ {\rm meV}$.  With the
parameters given above this means that we must keep a total of 134
bulk Landau levels.
Using the criteria stated above, we need 350 lattice points
in a sample $30 l$ wide. A second impenetrable wall is placed at
$x = 30 l$ to aid in normalization, but we ignore any nucleation at that
surface. The finite difference version of the BdG equations
is then an $700\times 700$ matrix equation.
The range of $x_{0}$ is not limited to lie within
the sample; indeed, the states with $x_{0}<0$ are precisely the edge states.
We use $800$ orbit centers ranging from $x_{0} = -35 l$ to $35 l$,
which ensures that all of the states in question have attained their bulk
values. Convergence was reached after five iterations, requiring ten hours on
an
IBM RS 6000/370  workstation.

\paragraph*{Discussion of Results.}
Our result for the amplitude of the self-consistent pair potential is given in
Fig.\ \ref{fig1}; this result is plotted against the GL theory result,
taken from Ref.\ \cite{fink65}.  The pair potential does indeed vanish at the
material surface and exhibits large Friedel oscillations away from the
surface,  with a period which is approximately $\pi/k_{F}=34\ {\rm \AA}$.
Such oscillations of the pair potential near a surface also occur in zero field
\cite{stojkovic93,troy},  and in the vicinity of a magnetic impurity
\cite{kummel72}, and are the result of pair-breaking by the surface and
impurity.
The self-consistent orbit center for the pair potential is $X_0 = 0.19 l$,
which is close to the position of the first maximum of the amplitude
of the pair potential.  The existence of a surface sheath at these
relatively high magnetic fields is not inconsistent with variational
calculations  at $T=0$ \cite{kulik69,hu69}, which give
$H_{c3}(0) \ge 1.95 H_{c2}(0)$.  We have repeated our calculations
at a field of 4.5 T, and find that the maximum amplitude of the
pair potential is  decreased.
Likewise, increasing the temperature results in a smaller
amplitude.   We have not been able to pin down
$H_{c3}(T)$ using our method, as achieving self-consistency becomes
delicate when the pair potential is small.  The phase
boundary is best determined by direct numerical
solution of the linearized gap equation \cite{chungyu}, which will also
us to attack the question of re-entrant behavior in $H_{c3}$.

The resulting change in the electronic structure can be seen in the
thermally broadened local density of states, $N(x,E)$, which we have
plotted at constant
energy, Fig.\ \ref{fig2}, and constant position,  Fig.\ \ref{fig3}.
In Fig.\ \ref{fig2} we see that at low energies the wavefunctions
have a reduced amplitude in the vicinity of the maximum of the pair
potential.  Due to the presence of local superconducting order,
in Fig.\ \ref{fig3} we see that close to the surface
there is a suppression in the density of states at low energies,
with a corresponding enhancement at energies above the
bulk gap.  Farther from the surface we obtain the bulk normal
density of states.  We expect that this structure could be resolved
by a STM measurement which would scan from the interior of a sample
to a surface parallel to the applied magnetic field.  Such a
measurement would provide the first direct observation of the
superconducting surface sheath.

We would like to thank Drs. S. Girvin, A. MacDonald, C.-Y. Mou, and J. D. Shore
 for useful discussions, and Dr. W. Pesch for bringing Ref.\ \cite{scotto92}
to our attention.  This work was supported by NSF Grant No. DMR 92-23586
and Jeffress Trust Grant J-289. A. T. D. gratefully acknowledges an
Alfred P. Sloan Foundation Fellowship.


\begin{figure}
\caption{The pair potential amplitude calculated from the BdG
equations,  compared with the GL result \protect\cite{fink65},
at $H/H_{c2}(T) = 1.83$ and $T/T_c = 0.28$.
The BdG solution vanishes at the surface and exhibits
strong Friedel oscillations as a result of the impenetrable wall at $x=0$.}
\label{fig1}
\end{figure}

\begin{figure}
\caption{Thermally averaged local density of states for as a function of
position (normalized to the normal local
density of states), at  $H/H_{c2}(T) = 1.83$ and $T/T_c = 0.28$.
For $E=0$ the electron states have been pushed away from the material
surface by the nucleation of the pair potential. At higher
energies ($E=1.1\ {\rm meV}$) the effect becomes smaller, and vanishes
altogether at very high energies.}
\label{fig2}
\end{figure}

\begin{figure}
\caption{Thermally averaged local density of states
as a function of energy (normalized to the normal local density of states),
at  $H/H_{c2}(T) = 1.83$ and $T/T_c = 0.28$.
Close to the surface ($x=0.1, 0.5$) there is a significant depletion
in the LDOS at low energies, as well as an enhancement at energies
above the bulk gap of $\Delta(0) = 1.1\ {\rm meV}$.  When $x=1.5$
the LDOS approaches its bulk normal state value.}
\label{fig3}
\end{figure}

\end{document}